\documentclass[10.5pt,a4paper]{article}
\usepackage[dvipdfmx]{graphicx}
\usepackage{subfigure}
\usepackage{here}
\graphicspath{{picture/}} 
\usepackage{comment}
\usepackage{enumerate}
\usepackage{paralist}
\usepackage{color}
\usepackage{ulem}
\usepackage{cancel}
\usepackage{setspace}  
\usepackage{lineno}
\usepackage{graphicx}
\usepackage[utf8]{inputenc}
\DeclareUnicodeCharacter{3000}{\ } 

\makeatletter
\def\Hline{\noalign{\ifnum0=`}\fi\hrule\@height 3.\arrayrulewidth \futurelet\reserved@a\@xhline}
\makeatother

\usepackage[top=30truemm,bottom=30truemm,left=20truemm,right=20truemm]{geometry}
\usepackage{fancyhdr}
\usepackage{amsmath}
\usepackage{amssymb}
\usepackage{mathcomp}
\usepackage{cases}
\usepackage{bm}

\usepackage{authblk} 
\usepackage{comment}

\title{Sagnac-type neutron displacement-noise-free interferometeric gravitational-wave detector}

\author[1,$\dagger$]{Yuki Kawasaki}
\author[1]{Shoki Iwaguchi}
\author[1]{Tomohiro Ishikawa}
\author[2, 3]{Atsushi Nishizawa}
\author[4,1]{Masaaki Kitaguchi}
\author[5]{Yutaka Yamagata}
\author[6]{Yanbei Chen}
\author[1]{Bin Wu}
\author[1]{Ryuma Shimizu}
\author[1]{Kurumi Umemura}
\author[1]{Kenji Tsuji}
\author[1,4]{Hirohiko Shimizu}
\author[7,2]{Yuta Michimura}
\author[8]{Kazuhiro Kobayashi}
\author[8]{Takafumi Onishi}
\author[1,4]{Seiji Kawamura}

\affil[1]{Department of Physics, Nagoya University, Nagoya, Aichi 464-8602, Japan;\\
\mbox{Correspondence: kawasaki\_y@u.phys.nagoya-u.ac.jp; Tel.:+81-52-789-6194}}
\affil[2]{Department of Physics, Graduate School of Advanced Science and Engineering, Hiroshima University, 1-3-1 Kagamiyama, Higashi-Hiroshima, Hiroshima 739-8526, Japan;}
\affil[3]{Research Center for the Early Universe (RESCEU), Graduate School of Science, The University of Tokyo, Tokyo 113-0033, Japan;}
\affil[4]{The Kobayashi-Maskawa Institute for the Origin of Particles and the Universe, Nagoya University, Nagoya, Aichi 464-8602, Japan;}
\affil[5]{RIKEN Center for Advance Photonics, RIKEN, Hirosawa 2-1, Wako, Saitama 351-0198, Japan;}
\affil[6]{Theoretical Astrophysics 350-17, California Institute of Technology, Pasadena, California 91125, USA;}
\affil[7]{LIGO Laboratory, California Institute of Technology, Pasadena, California 91125, USA;}
\affil[8]{Equipment Development Support Section, Technical Center of Nagoya University, Nagoya, Aichi 464-8602, Japan;}

\begin{document}
\maketitle

\vspace{1.0\baselineskip}
\begin{abstract}
\noindent
The detection of low-frequency gravitational waves on Earth requires the reduction of displacement noise, which dominates the low-frequency band. One method to cancel test mass displacement noise is a neutron displacement-noise-free interferometer (DFI). This paper proposes a new neutron DFI configuration, a Sagnac-type neutron DFI, which uses a Sagnac interferometer in place of the Mach-Zehnder interferometer. We demonstrate that a sensitivity of the Sagnac-type neutron DFI is higher than that of a conventional neutron DFI with the same interferometer scale. This configuration is particularly significant for neutron DFIs with limited space for construction and limited flux from available neutron sources.
\end{abstract}

\newpage
\section{Introduction}
\begin{figure}[H]
    \centering
    \includegraphics[clip,width=8.0cm]{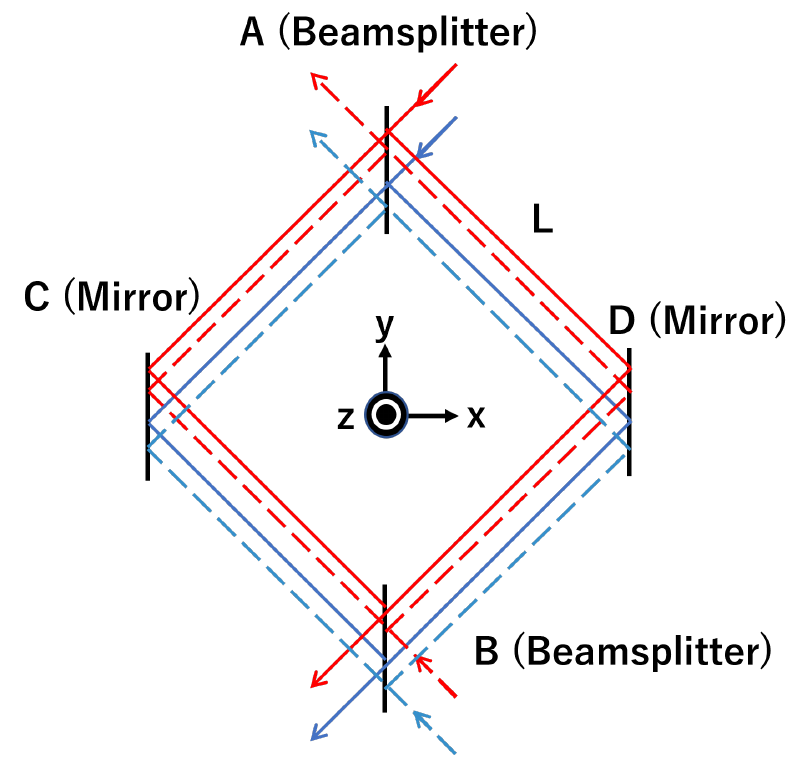}
    \caption{Configuration of single two-velocity neutron DFI as viewed from the z-axis direction. The black bars represent the test masses, mirrors and beamsplitters. The red and blue arrows show the trajectories of the neutrons with incident velocity $v_1$ and $v_2$ projected into horizomtal plane, respectively. The solid arrows are the trajectories of neutrons incident through beamsplitter (BS) A, while the dashed arrows represent the trajectories of neutrons incident through BS B. Mirrors C and D are located on the x-axis and BSs A and B are located on the y-axis. The origin in the xy-plane represents the center of the square formed by the neutron orbits.}
    \label{Single two-velocity}
\end{figure}

Since 2015, when LIGO first detected gravitational waves (GWs), numerous GW events have been observed \cite{ref1, ref2, ref3}. The frequencies of GWs depend on their sources. For example, GWs in the low-frequency band around 1 Hz include important physical phenomena such as ``primordial gravitational waves (PGWs)" \cite{ref4}. However, the sensitivity of GW detectors in the low frequency band is limited by displacement noise. This noise arises from the fluctuating displacement of the detector's test masses, due to thermal noise, noise from the suspension system, and seismic noise. One way to minimize this displacement noise is to send GW detectors up into space, as for LISA \cite{ref5, ref6} and DECIGO \cite{ref7, ref8}. Space-based GW detectors can reduce some displacement noise, such as suspension system noise, and ground vibration noise by floating the test mass in space. Cooling mirrors and suspension systems in space can also reduce thermal noise. However, space-based detectors require a significant amount of time, effort, and cost. The sensitivity in the low-frequency band also is limited by radiation pressure noise, which is one of the sources of displacement noise. Thus, reducing displacement noise is also essential.
Displacement-noise-free interferometery (DFI) is one of the methods for canceling all displacement noise \cite{ref9}. In a coordinate system in the transverse-traceless gauge, GW signals and displacement noise can be distinguished \cite{ref10, ref11}. Therefore, multiple interferometer signals can be used to cancel displacement noise while preserving GW signals \cite{ref12, ref13}. This is the principle behind a DFI. \\
DFI is most sensitive in a limited frequency band that corresponds to the inverse of the time required for the beam to propagate between test masses. For example, a DFI with several kilometers of arm length is mostly sensitive near $10^5\,\mathrm{Hz}$, which is much higher than the 1\,Hz range where displacement noise is dominant. This is because of the high propagation speed of the laser light. Therefore, the neutron DFI was devised, in which a neutron beam is injected into the DFI. Neutrons have a finite mass, and their speed is much slower than that of photons. Therefore, even a DFI with several kilometers of arm length has good sensitivity around 1 Hz. Furthermore, the ability to select the speed of the incident neutrons is an advantage for optimizing the sensitivity of a DFI. \\
Various configurations of neutron DFI have already been developed \cite{ref14, ref15, ref16}. In this paper, we present a neutron DFI configuration that uses a Sagnac-type interferometer. In the conventional configuration (single, two-velocity), shown in Figure\,\ref{Single two-velocity}, a single Mach-Zehnder interferometer is injected with neutrons of two velocities from two relative directions \cite{ref15}. The solid arrows are the trajectories of neutrons incident through beamsplitter (BS) A, while the dashed arrows represent the trajectories of neutrons incident through BS B. After entering the Mach-Zehnder interferometer, the neutrons are divided into right and left paths by a beamsplitter (BS) and propagate along the two sides of the interferometer under the influence of GWs. The photodetector (PD) observes the interference state of the neutrons in the right and left paths. A Sagnac-type neutron DFI is injected with neutrons of four velocities from one direction. The PD observes the interference state of clockwise- and counterclockwise-propagating neutrons. This configuration allows a Sagnac-type neutron DFI to achieve higher sensitivity than a conventional neutron DFI of the same size interferometer and with the same number of neutrons.\\
In this paper, we discuss the Sagnac-type neutron DFI in terms of configuration and sensitivity. We discuss the configuration in Section\,\ref{Configuration of Sagnac-type neutron DFI}, the method of canceling displacement noise in Section\,\ref{Method of cancelling displacement noise}, the GW signals from neutron DFIs in Section\,\ref{Gravitational wave signal of neutron DFI}, noise and sensitivity comparisons with the single, two-velocity neutron DFI in Section\,\ref{Sensitivity}, and consider characteristics of the sensitivity curve in Section\,\ref{Discussion}. Finally we present conclusions in Section\,\ref{Conclusion}.

\nolinenumbers
\begin{figure}[H]
    \centering
    \includegraphics[clip,width=8.0cm]{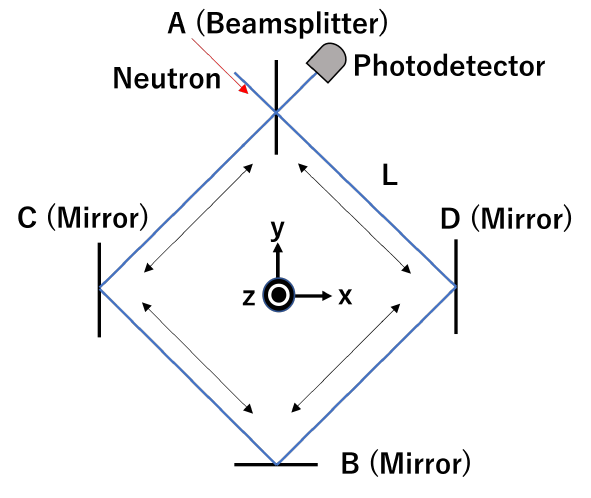}
    \caption{Configuration of a Sagnac-type neutron DFI viewed along the z-axis. The black bars represent test masses, mirrors and beamsplitter. The blue lines show the trajectories of the neutrons. Mirrors C and D are placed on the x-axis, while BS A and mirror B are placed on the y-axis. The origin in the xy-plane is the center of the square drawn by the neutron orbit. The origin on the z-axis is the height at which a neutron trajectory intersects the plane of BS A.}
    \label{Sagnac_xy}
\end{figure}
\section{Configuration of a Sagnac-type neutron DFI}
\label{Configuration of Sagnac-type neutron DFI}

\begin{figure}[H]
    \centering
    \includegraphics[clip,width=12.0cm]{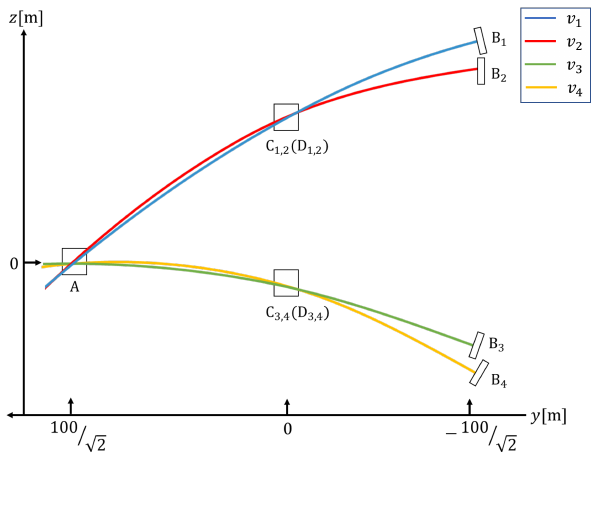}
    \caption{Pathway of neutrons in the yz plane. Rectangles and squares indicate mirrors and beamsplitters, respectively. Mirrors located in the same position are the same, while mirrors located in different positions are distinct. Mirrors $\mathrm{B}_i$ (i = 1 to 4) are angled in the yz-plane to be perpendicular to the trajectories of $n_i$. The height at which all neutrons hit the BS A is $0\,\mathrm{m}$.}
    \label{Sagnac_yz}
\end{figure}

The configuration of a Sagnac-type interferometer is shown in Figure\,\ref{Sagnac_xy} from a bird's-eye view (xy plane) and in Figure\,\ref{Sagnac_yz} from the side view (yz plane). Gravity acts in the z direction.The propagation of the neutron trajectory in the xy plane is a square with a side length of $L$. BS A and mirrors B, C, and D are placed at the vertices of the square. Neutrons incident on A are divided into two directions and propagate in the interferometer in clockwise and counterclockwise orbits.
Neutrons with four different velocities $v_i$ ($i=1, 2, 3, 4$) are injected into the Sagnac-type neutron DFI. The horizontal and vertical velocities of neutrons with incident velocity $v_i$ ($n_i$) are denoted by $v_{i}^{\mathrm{h}}$ and $v_{i}^{\mathrm{v}} $, respectively. 
As shown in Figure\,\ref{Sagnac_yz}, $n_i$ hit mirrors $\mathrm{C}_i$ and $\mathrm{D}_i$ after a time $T_i(=L/v_{i} ^{\mathrm{h}})$ from the incident, and at mirror $\mathrm{B}_i$ after another $T_i$. Mirror $\mathrm{B}_i$ is angled in the xz plane so that it is perpendicular to the trajectory of $n_i$ in the yz plane. Then, only the x-component of the neutron velocity is preserved, and the neutron is reflected back. After reflection at mirror $\mathrm{B}_i$, both clockwise and counterclockwise orbits are symmetrical about the y-axis. Then, $2T_i$ after the reflection at mirror $\mathrm{B}_i$ the clockwise neutrons pass through 
BS A, and the counterclockwise neutrons are reflected at BS A. Their interference states are observed by the PD. In this paper, we set $t=0$ to be the time when the clockwise and counterclockwise neutrons are simultaneously reflected by mirror $\mathrm{B}_i$. Thus, neutrons are reflected by mirror $\mathrm{C}_i$ and $\mathrm{D}_i$ at $t = \pm T_i$, and hit BS A at $t = \pm 2T_i$. This means that neutrons with different velocities are reflected simultaneously only at mirror $\mathrm{B}_i$ and hit mirrors $\mathrm{C}_i$ and $\mathrm{D}_i$ and BS A at different times.

\subsection{Neutron trajectory}

Let us consider the trajectory of $n_i$ taking into account the effect of falling due to gravity. Here, we ignore the mirror displacement temporarily and focus only on the neutrons trajectories. The coordinate at which the neutrons collide with A is denoted by

\begin{equation}
\label{x_A}
    \bm{x}_{i \mathrm{A}} = (0, \frac{L}{\sqrt{2}}, 0).
\end{equation}
The velocity of the neutrons incident on A is denoted by

\begin{equation}
\label{v_A}
    \bm{v}_{i} = (\frac{1}{\sqrt{2}}v_{i}^{\mathrm{h}}, -\frac{1}{\sqrt{2}}v_{i}^{\mathrm{h}}, v_{i}^{\mathrm{v}}).
\end{equation}
After a time interval $T_i$ elapses after the neutrons' incidence on A, the neutrons collide with mirrors C and D. The velocities of the neutrons as they propagate from A to C and from A to D are given by

\begin{align}
\label{v_AC}
    \bm{v}_{i\mathrm{AC}}(t) &= (-\frac{1}{\sqrt{2}}v_{i} ^{\mathrm{h}}, -\frac{1}{\sqrt{2}}v_{i} ^{\mathrm{h}}, v_{i} ^{\mathrm{v}} - g(t+2T_i)) ~(-2 T_i\leq t\leq -T_i),\\
    \bm{v}_{i\mathrm{AD}}(t) &= ( \frac{1}{\sqrt{2}}v_{i} ^{\mathrm{h}}, -\frac{1}{\sqrt{2}}v_{i} ^{\mathrm{h}}, v_{i} ^{\mathrm{v}} - g(t+2T_i)) ~(-2T_i\leq t\leq -T_i).
\end{align}
The coordinates at where the neutrons impact mirrors C and D are

\begin{align}
\label{x_CD}
    \bm{x}_{i\mathrm{C}} &= ( -\frac{L}{\sqrt{2}}, 0, v_{i} ^{\mathrm{v}} T_i - \frac{1}{2}g T^{2}_i),\\
    \bm{x}_{i\mathrm{D}} &= (  \frac{L}{\sqrt{2}}, 0, v_{i} ^{\mathrm{v}} T_i - \frac{1}{2}g T^{2}_i).
\end{align}
The velocities of the neutrons as they propagate from C to B and from D to B and the coordinate where they impact mirror B are given by

\begin{align}
\label{v_CB}
    \bm{v}_{i\mathrm{CB}}(t) &= (\frac{1}{\sqrt{2}}v_{i} ^{\mathrm{h}}, -\frac{1}{\sqrt{2}}v_{i} ^{\mathrm{h}}, v_{i} ^{\mathrm{v}} - g(t+2T_i))~(-T_i< t\leq 0),\\
    \bm{v}_{i\mathrm{DB}}(t) &= (-\frac{1}{\sqrt{2}}v_{i} ^{\mathrm{h}}, -\frac{1}{\sqrt{2}}v_{i} ^{\mathrm{h}}, v_{i} ^{\mathrm{v}} - g(t+2T_i))~(-T_i< t\leq 0),\\
    \bm{x}_{i\mathrm{B}} &= (0, -\frac{L}{\sqrt{2}}, 2v_{i} ^{\mathrm{v}} T_i - \frac{1}{2}g (2T_i)^{2}).
\end{align}
After reflection at mirror B, oriented in the yz plane to be perpendicular to the trajectories of the neutrons, the clockwise-propagating neutrons propagate from B to C and from C to A. The counterclockwise-propagating neutrons propagate from B to D and from D to A. The velocities of the neutrons are given by

\begin{align}
    \bm{v}_{i\mathrm{BD}}(t) &=  (\frac{1}{\sqrt{2}}v_{i} ^{\mathrm{h}},  \frac{1}{\sqrt{2}}v_{i} ^{\mathrm{h}}, -v_{i} ^{\mathrm{v}} - g(t-2T_i))~(0< t\leq T_i),\\
    \bm{v}_{i\mathrm{DA}}(t) &=  (-\frac{1}{\sqrt{2}}v_{i} ^{\mathrm{h}},  \frac{1}{\sqrt{2}}v_{i} ^{\mathrm{h}}, -v_{i} ^{\mathrm{v}} - g(t-2T_i))~(T_i < t\leq 2T_i),\\
    \bm{v}_{i\mathrm{BC}}(t) &=  (-\frac{1}{\sqrt{2}}v_{i} ^{\mathrm{h}},  \frac{1}{\sqrt{2}}v_{i} ^{\mathrm{h}}, -v_{i} ^{\mathrm{v}} - g(t-2T_i))~(0< t\leq T_i),\\
    \bm{v}_{i\mathrm{CA}}(t) &=  (\frac{1}{\sqrt{2}}v_{i} ^{\mathrm{h}},  \frac{1}{\sqrt{2}}v_{i} ^{\mathrm{h}}, -v_{i} ^{\mathrm{v}}  -g(t-2T_i))~(T_i <t\leq 2T_i).
\end{align}
Mirror $\mathrm{B}_i$ is rotated about the x-axis to be perpendicular to the trajectory of neutrons projected on the yz plane. Thus, the z component of the velocity of the neutrons reverses its sign after reflection by mirror $\mathrm{B}_i$.
Neutrons $n_i$ propagate through this trajectory. The velocities of clockwise- and counterclockwise-propagating neutrons $\bm{v}_i^{l}$ $(l = \mathrm{c, cc})$ can be defined as 

\begin{equation}
    \bm{v}_i^{\mathrm{c}}(t) = \left\{
    \begin{array}{l}
    \bm{v}_{i\mathrm{AD}}(t)~(-2T_i\leq t\leq -T_i)\\
    \bm{v}_{i\mathrm{DB}}(t)~(-T_i< t\leq 0)\\
    \bm{v}_{i\mathrm{BC}}(t)~(0< t\leq T_i)\\
    \bm{v}_{i\mathrm{CA}}(t),~(T_i <t\leq 2T_i)\\
    \end{array}\right.
\end{equation}     

\begin{equation}
    \bm{v}_i^{\mathrm{cc}}(t) = \left\{
    \begin{array}{l}
     \bm{v}_{i\mathrm{AC}}(t)~(-2T_i\leq t\leq -T_i)\\
     \bm{v}_{i\mathrm{CB}}(t)~(-T_i< t\leq 0)\\
     \bm{v}_{i\mathrm{BD}}(t)~(0< t\leq T_i)\\
     \bm{v}_{i\mathrm{DA}}(t),~(T_i <t\leq 2T_i)\\
    \end{array}\right.
\end{equation}   

\subsection{Neutron pathway for DFI realization}
\begin{figure}[H]
    \centering
    \includegraphics[clip,width=12.0cm]{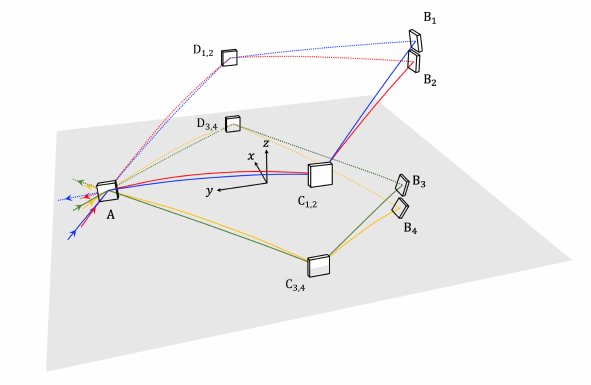}
    \caption{Neutron pathways for complete displacement noise cancelation. Each colored line corresponds to the trajectory of one group of neutrons, $n_i$. All neutrons enter the interferometer through the same point on A. The neutrons represented by the red and blue lines impact the same point, $\mathrm{C}_{1,2}$, while the neutrons represented by the yellow and green lines impact the same points, $\mathrm{C}_{3,4}$. All neutrons are then reflected by the mirrors $\mathrm{B}_i$ and follow trajectories that are symmetric about the y-axis.}
    \label{Sagnac_path}
\end{figure}

The realization DFI necessitates combining pairs of neutron signals that contain the displacement noise from the same test mass. This is because displacement noise from different points, even on the same test mass, are not correlated. To cancel the displacement noise at a point on C(D), two neutron signals that hit the same point on C(D) are required. Similarly, to cancel the displacement noise at a point on A, we need two signals that reflect at the same point on A, with the displacement noise at a point on C(D) already canceled.
To summarize, $n_1$ and $n_2$ and $n_3$ and $n_4$ must impact the same point on C(D), and all four neutrons must impact the same point on A. We determine $v_{i}^{\mathrm{h}}$ and $v_{i}^{\mathrm{v}}$ for each neutron group, $n_i$, to satisfy these conditions using the following procedure:\\
1. Set four horizontal incident neutron velocities $v_i^{\mathrm{h}}$ ($i = 1 ~\text{to}~ 4$).\\
2. Set $v_2^{\mathrm{h}}$ as the slowest horizontal velocity and $v_4^{\mathrm{h}}$ as the fastest horizontal velocity.\\
3. Determine $v_{2}^{\mathrm{v}}, v_{4}^{\mathrm{v}}$ such that $v_2$ and $v_4$ are within the range of neutron velocities that can be injected. In this paper, we set the slowest neutron velocity as 75\,m/s and the fastest neutron velocity as 100\,m/s. (This is the range of neutron velocities for which comparable fluxes can be expected.) \\
4. Calculate  $z_{\mathrm{C} i}$, the height at which $n_i$ impact mirror $\mathrm{C}_i$ ($\mathrm{D}_i$) as 

\begin{equation}
\label{Z_C}
    z_{\mathrm{C} i} = v_{i} ^{\mathrm{v}}T_i - \frac{1}{2}gT_i^2, 
\end{equation}
and determine $v_{1}^{\mathrm{v}}$ and $v_{3}^ {\mathrm{v}}$ to satisfy $z_{\mathrm{C} 1} = z_{\mathrm{C} 2}$ and $z_{\mathrm{C} 3} = z_{\mathrm{C} 4}$. Note that the height at which all neutrons impact mirror A is $0\,\mathrm{m}$.
Figure\,\ref{Sagnac_path} shows the neutron pathways that satisfy the conditions for the neutron velocities.

\section{Method of canceling displacement noise}
\label{Method of cancelling displacement noise}
\subsection{Displacement noise in the time domain}
First, let us consider only the displacement noise experienced by neutrons, $n_i$. When considering the displacement noise that the neutrons receive from mirrors $\mathrm{B}_i$, $\mathrm{C}_i$, and $\mathrm{D}_i$, we omit the subscripts for the mirrors B, C, and D. We set $\bm{d}_j$ as the displacement of test mass $j$ ~($j =$A, B, C, D) in the direction normal to its surface. The magnitude of the phase change of the neutrons, $\phi_{i j}^{\mathrm{dis}}(t)$ caused by $\bm{d}_j$ is expressed in terms of the wavenumber of the neutrons, $\bm{k}_i$ as

\begin{equation}
\label{phi_ij}
    \phi_{i j}^{\mathrm{dis}}(t)= 2\bm{k}_i^{l}(t) \cdot \bm{d}_j(t), ~~\bm{k}_i^l = \frac{m}{\hbar} \bm{v}_i^{l}(t).
\end{equation}
Here, $m$ is the mass of the neutrons, and $\hbar$ is Dirac's constant.
Table\,\ref{displacementnoise} shows the displacement noise that the clockwise and counterclockwise-propagating neutrons experience. Note that $t=0$ corresponds to the time when the neutrons are simultaneously reflected by mirror B. Also, since clockwise-propagating neutrons traverse BS A twice, they do not receive displacement noise from BS A.

\begin{table}[hbtp]
  \caption{Phase changes that the neutrons groups, $n_i$, receive.}
  \label{displacementnoise}
  \centering
  \begin{tabular}{|c||c|c|c|c|c|}
    \hline
    Test mass  & A & C & B & D & A \\
    \hline
    clockwise $\phi_{i} ^{\mathrm{c},\mathrm{dis}}$    &            0         & $\phi_{i \mathrm{C}}^{\mathrm{dis}}(T_i)$  & $\phi_{i \mathrm{B}}^{\mathrm{dis}}(0)$ & $\phi_{i \mathrm{D}}^{\mathrm{dis}}(-T_i)$ & 0 \\
    counterclockwise $\phi_{i} ^{\mathrm{cc},\mathrm{dis}}$     & $\phi_{i \mathrm{A}}^{\mathrm{dis}}(-2T_i)$ & $\phi_{i \mathrm{C}}^{\mathrm{dis}}(-T_i)$ & $\phi_{i \mathrm{B}}^{\mathrm{dis}}(0)$ & $\phi_{i \mathrm{D}}^{\mathrm{dis}}(T_i)$ & $\phi_{i \mathrm{A}}^{\mathrm{dis}}(2T_i)$  \\
    \hline
    \end{tabular}
\end{table}
The total phase change of clockwise- and counterclockwise-propagating $n_i$ is given by

\begin{equation}
\label{phi_ir}
    \phi_{i} ^{\mathrm{c},\mathrm{dis}} = \phi_{i \mathrm{C}}^{\mathrm{dis}}(T_i) + \phi_{i \mathrm{B}}^{\mathrm{dis}}(0) + \phi_{i \mathrm{D}}^{\mathrm{dis}}(-T_i),
\end{equation}

\begin{equation}
\label{phi_il}
    \phi_{i} ^{\mathrm{cc},\mathrm{dis}} = \phi_{i \mathrm{A}}^{\mathrm{dis}}(-2T_i) + \phi_{i \mathrm{C}}^{\mathrm{dis}}(-T_i) + \phi_{i \mathrm{B}}^{\mathrm{dis}}(0) + \phi_{i \mathrm{D}}^{\mathrm{dis}}(T_i) + \phi_{i \mathrm{A}}^{\mathrm{dis}}(2T_i).
\end{equation}
The phase change of the interfered neutrons at the PD is given by

\begin{align}
\label{dis_in_time}
   \phi^{\mathrm{dis}}_i &\equiv \phi_{i} ^{\mathrm{c}} - \phi_{i} ^{\mathrm{cc}}\\ \nonumber
    &= -[\phi^{\mathrm{dis}}_{i \mathrm{A}}(2T_i) + \phi^{\mathrm{dis}}_{i \mathrm{A}}(-2T_i)] + [\phi^{\mathrm{dis}}_{i \mathrm{C}}(T_i) -\phi^{\mathrm{dis}}_{i \mathrm{C}}(-T_i)] + [\phi^{\mathrm{dis}}_{i \mathrm{D}}(-T_i) -\phi^{\mathrm{dis}}_{i \mathrm{D}}(T_i)].
\end{align}
The clockwise- and counterclockwise-propagating neutrons impact the same point on mirror B simultaneously, thus they receive identical displacement noise from the motion of B. As a result, the signal $\phi_i$ after they interfere contains no displacement noise from B. However, due to the fact that the clockwise- and counterclockwise-propagating neutrons interact with A, C, and D at different times, displacement noise cannot be fully eliminated.
\subsection{Displacement noise in frequency domain}
In order to cancel the displacement noise of A, C, and D, we must consider the displacement noise in the frequency domain. In this paper, we define the Fourier transform as follows

\begin{equation}
     f(t) \equiv \int F(\omega)e^{-i\omega t}d\omega.
\end{equation}
Note that  lowercase letters represent variables in the time domain and uppercase letters represent variables in the frequency domain. We can express $\bm{d}_j(t)$ as 

\begin{equation}
   \bm{d}_j(t) \equiv \int \bm{D}_j(\omega)e^{-i\omega t} d\omega,
\end{equation}
where $\bm{D}_j(\omega)$ is the complex amplitude of the mirror displacement at frequency $\omega$. We can express $\phi_{i j}^{\mathrm{dis}}(t)$ in the frequency domain as 

\begin{equation}
\label{Phi(omega)}
    \Phi_{i j}^{\mathrm{dis}}(\omega) = 2\bm{k}_i \cdot \bm{D}_j(\omega).
\end{equation}
Since the normals of the surfaces of test masses A, C, and D have only x-components, Equation\,\ref{phi_ij} can be rewritten using the unit vector in the x direction $\bm{e_x}$ as

\begin{align}
    \phi_{i j}^{\mathrm{dis}}(t)= \frac{2m}{\hbar}(\bm{v}_i(t) \cdot \bm{e_x}) (\bm{d}_j(t) \cdot \bm{e_x}) = \frac{2mv_{ix}}{\hbar} d_j(t),
\end{align}
where $v_{ix}$ is the x component of $\bm{v}_i$. Although the z component of $\bm{k}_i$ changes with time due to gravity, it does not affect $\phi_{i j}^{\mathrm{dis}} (t)$, which is proportional to the inner product of $\bm{k}_i$ and $\bm{D}_j$.
We treat $\bm{k}_i$ as a time-invariant variable because the calculation of $\Phi_{ij}(\omega)$ (in Equation\,\ref{Phi(omega)}) does not require consideration of the time variation of $\bm{k}_i$.
Considering the velocity of the neutrons at the time they impact each test mass, the $\omega$ component of Equation\,\ref{dis_in_time} is given by

\begin{align}
\begin{split}
 \Phi^{\mathrm{dis}}_i(\omega) &= -2\bm{k}_i\cdot[ \bm{D}_{\mathrm{A}} (e^{-i2\omega T_i}-e^{i2\omega T_i})\\
                                         &+ \bm{D}_{\mathrm{C}} (e^{-i\omega T_i}-e^{i\omega T_i}) - \bm{D}_{\mathrm{D}} (e^{i\omega T_i}-e^{-i\omega T_i}) ].
\end{split}
\end{align}
\subsubsection{Method of cancelling displacement noise of mirrors C and D}
First, we consider the displacement noise of mirrors C and D in the frequency domain. Table\,\ref{displacementnoise of C and D} shows the displacement noise of C and D in the interferometer signals of the neutrons.

\begin{figure}[H]
    \centering
    \includegraphics[clip,width=15.0cm]{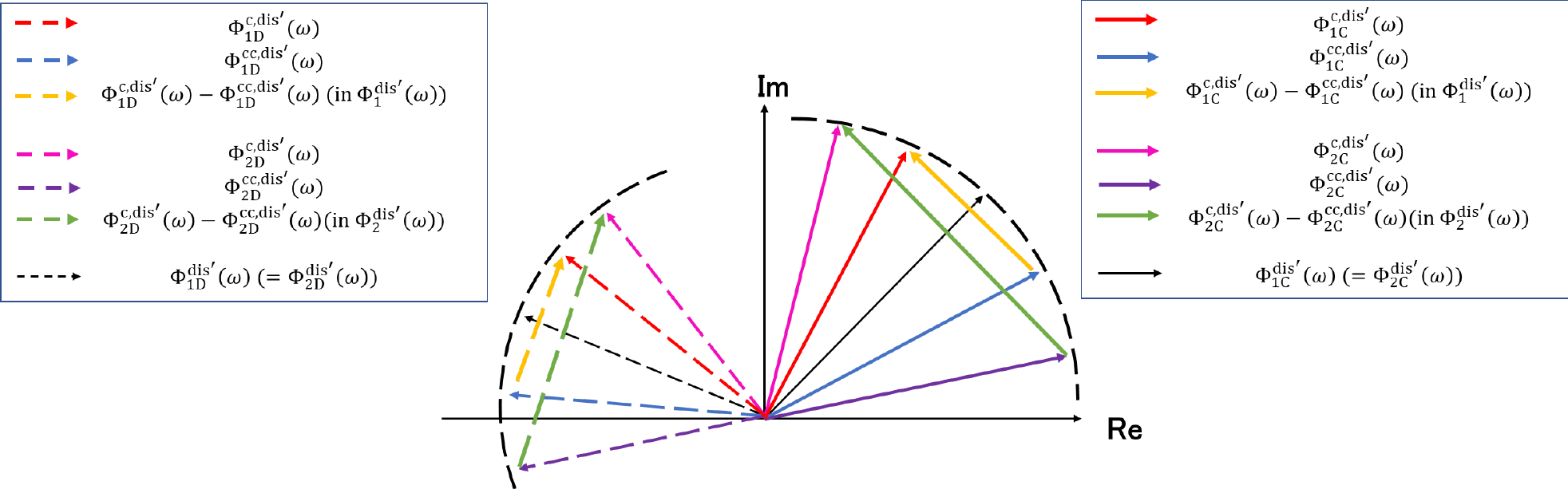}
    \caption{Phasor diagram that illustrates the cancelation of displacement noise from mirrors C and D. Each arrow represents the displacement noise that the neutrons receive with the length of the arrows indicating the amplitude of the displacement noise, and  the direction indicating the phase of the displacement noise. The solid arrow represents the displacement noise of mirror C, and the dashed arrow represents the displacement noise of mirror D. The color of the arrows correspond to the source of the displacement noise. The red line represents the displacement noise of mirror C or D in $\Phi_{1} ^{\mathrm{c},\mathrm{dis}}(\omega)$. The blue line represents the displacement noise in $\Phi_{1} ^{\mathrm{c},\mathrm{dis}}(\omega)$, while the pink and purple lines represent the displacement noise in $\Phi_{2}^{\mathrm{c},\mathrm{dis}}(\omega)$, respectively. The green and yellow lines represent the displacement noise in $\Phi_1^{\mathrm{dis}}(\omega)$ and $\Phi_2^{\mathrm{dis}}(\omega)$, respectively. Finally, the black line represents the displacement noise at $t=0$.}
    \label{PD_CD}
\end{figure}

\begin{table}
  \caption{Displacement noise of the test masses C and D}
  \label{displacementnoise of C and D}
  \centering
  \begin{tabular}{|l||c|c|}
    \hline
    Test masses  & C & D \\
    \hline
    clockwise     $\Phi_{i} ^{\mathrm{c}}$    & $2\bm{k}_i \cdot \bm{D}_\mathrm{C}(\omega)e^{-i\omega T_i}$ & $2\bm{k}_i \cdot \bm{D}_\mathrm{D}(\omega)e^{i\omega T_i}$  \\
    counterclockwise     $\Phi_{i} ^{\mathrm{cc}}$    & $2\bm{k}_i \cdot \bm{D}_\mathrm{C}(\omega)e^{i\omega T_i}$   & $2\bm{k}_i \cdot \bm{D}_\mathrm{D}(\omega)e^{-i\omega T_i}$ \\
    Interferometer Signal $\Phi_i$    & $2\bm{k}_i \cdot \bm{D}_\mathrm{C}(\omega)(e^{-i\omega T_i} - e^{i\omega T_i})$       & $2\bm{k}_i \cdot \bm{D}_\mathrm{D}(\omega)(e^{i\omega T_i} - e^{-i\omega T_i})$ \\
    \hline
    \end{tabular}
\end{table}
$\Phi_i^{\mathrm{dis}}(\omega)$ contains the displacement noise of mirrors C and D. 
In order to normalize the interferometer signals with different neutron velocities, we divide $\Phi^{\mathrm{dis}}_i(\omega)$ by $v_{i}^{\mathrm{h}}/\sqrt{2}c$ and we obtain

\begin{align}
    \Phi^{\mathrm{dis}'}_i(\omega) \equiv \frac{\sqrt{2}c}{v_{i}^{\mathrm{h}}}\Phi^{\mathrm{dis}}_i(\omega) &= -\frac{4mc}{\hbar}[ D_{\mathrm{A}}(\omega)(-i\sin{2\omega T_i}) + D_{\mathrm{C}}(\omega)(-i\sin{\omega T_i}) + D_{\mathrm{D}}(\omega)(-i\sin{\omega T_i})]. 
\end{align}
Here, we take into account $n_1$ and $n_2$.
Since the ratio of the magnitudes of the displacement noise caused by C and D in ${\Phi^{\mathrm{dis}'}_1(\omega)}$ and ${\Phi^{\mathrm{dis}'}_2(\omega)}$ is $\sin{\omega T_1} : \sin{\omega T_2}$, they can be eliminated by the following signal processing:

\begin{gather}
\label{Cancel CD}
    \Phi^{\mathrm{dis}}_{12}(\omega) \equiv  \alpha_1 {\Phi^{\mathrm{dis}'}_1(\omega)} - \alpha_2 {\Phi^{\mathrm{dis}'}_2(\omega)},\\
    \alpha_1 = \sin{\omega T_2} ,\quad \alpha_2 = \sin{\omega T_1}.
\end{gather}
We can explain the fact that $\Phi_{12}^{\mathrm{dis}}(\omega)$ has no displacement noise from C and D with a phasor diagram  as shown in Figure\,\ref{PD_CD}.
The green and yellow lines represent the displacement noise in $\Phi_1^{\mathrm{dis}}(\omega)$ and $\Phi_2^{\mathrm{dis}}(\omega)$, respectively. Since the green and yellow lines (both solid and dashed arrows, which represent the displacement noise due to mirrors $\mathrm{C}_i$ and $\mathrm{D}_i$, respectively) are parallel in the phasor diagram, they can be simultaneously eliminated by subtracting a real coefficient.
The details of the calculations are as follows:

\begin{align}
\begin{split}
    \Phi^{\mathrm{dis}}_{12}(\omega) &= -\frac{4mc}{\hbar}[ D_{\mathrm{A}}(-i\sin{\omega T_2}\sin{2\omega T_1}) + (D_{\mathrm{C}} + D_{\mathrm{D}})(-i\sin{\omega T_1}\sin{\omega T_2})]\\
    &+\frac{4mc}{\hbar}[ D_{\mathrm{A}}(-i\sin{\omega T_1}\sin{2\omega T_2}) + (D_{\mathrm{C}} + D_{\mathrm{D}})(-i\sin{\omega T_1}\sin{\omega T_2})],
\end{split}\nonumber
    \\&= -i\frac{4mc}{\hbar}D_{\mathrm{A}}(\sin{\omega T_1}\sin{2\omega T_2}-\sin{\omega T_2}\sin{2\omega T_1}).
\end{align}
Similarly, we can obtain $\Phi^{\mathrm{dis}}_{34}(\omega)$, that has no displacement noise from C and D as 

\begin{align}
    \Phi^{\mathrm{dis}}_{34}(\omega) 
    &\equiv  \alpha_3 {\Phi^{\mathrm{dis}'}_3(\omega)} - \alpha_4 {\Phi^{\mathrm{dis}'}_4(\omega)},\nonumber\\
    &= -i\frac{4mc}{\hbar}D_{\mathrm{A}}(\sin{\omega T_3}\sin{2\omega T_4}-\sin{\omega T_4}\sin{2\omega T_3}).\\
    \alpha_3 &= \sin{\omega T_4} ,\quad \alpha_4 = \sin{\omega T_3}.
\end{align}

\subsubsection{Method of canceling displacement noise of A}
 In this subsection, we will focus on the displacement noise of BS A in the frequency domain. In the Sagnac-type neutron DFI configuration, the clockwise-propagating neutrons always pass through A and are thus not affected by its displacement. However, the counterclockwise neutrons accumulate the displacement noise of A twice, at $t=-2T_i$ and $t=2T_i$. If A is displaced by $\bm{d}_\mathrm{A}$ at $t=-2T_i$, the counterclockwise neutron propagation distance increases, leading to an advance in the neutron phase. Conversely, if A is displaced by $\bm{d}_\mathrm{A}$ at $t=2T_i$, the counterclockwise neutron propagation distance decreases, resulting in a delay in the neutron phase. As a result, the sign of the displacement noise of A is opposite for $t=-2T_i$ and $t=2T_i$.
The impact of the displacement noise of A on $\Phi_i^{\mathrm{dis}}(\omega)$ can be expressed as

\begin{equation}
    \Phi_i^{\mathrm{dis}}(\omega) = -2\bm{k}_i\cdot \bm{D}_\mathrm{A}(\omega)(e^{-i2\omega T_i} - e^{i2\omega T_i}).
\end{equation}
Similar to the method for canceling the displacement noise of mirrors C and D in Equation\,\ref{Cancel CD}, the displacement noise of A can be cancelled by combining two signals containing the displacement noise of A.
If we cancel the displacement noise of A using $\Phi_{12}^{\mathrm{dis}}(\omega)$ and $\Phi_{34}^{\mathrm{dis}}(\omega)$, which contain noise components from A only, we can obtain a signal $\Phi_{\mathrm{SagnacDFI}}(\omega)$ that has all displacement noise of the Sagnac-type DFI canceled:

\begin{equation}
    \Phi^{\mathrm{dis}}_{\mathrm{SagnacDFI}}(\omega) = \beta_1 \Phi_{12}^{\mathrm{dis}}(\omega) - \beta_2 \Phi_{34}^{\mathrm{dis}}(\omega) = 0,
\end{equation}
where

\begin{align}
    \beta_1 = \sin 2\omega T_3 \sin \omega T_4 - \sin 2\omega T_4 \sin \omega T_3\quad \mathrm{and} &\quad
    \beta_2 = \sin 2\omega T_1 \sin \omega T_2 - \sin 2\omega T_2 \sin \omega T_1.
\end{align}
Specifically, the coefficient of $\bm{D}_{\mathrm{A}}$ in $\Phi_{12}^{\mathrm{dis}}(\omega)$ ($=\beta_2$) and that in $\Phi_{34}^{\mathrm{dis}}(\omega)$ ($=\beta_1$) are multiplied by $\Phi_{34}^{\mathrm{dis}}(\omega)$ and $\Phi_{12}^{\mathrm{dis}}(\omega)$, respectively, and subtracted to cancel the BS A displacement noise.
\subsubsection{Sagnac effect in Sagnac-type neutron DFI}
In the Sagnac-type neutron DFI, neutrons undergo the Sagnac effect and experience a phase change when they are detected after circling the interferometer, if the interferometer rotates with a certain angular frequency $\omega_0$. From the perspective of the inertial system, the rotation of the interferometer can be expressed by the displacement of each test mass. However, in the DFI, all displacement noise is canceled. Thus, the signals obtained from a Sagnac-type neutron DFI,  $\Phi_{\mathrm{SagnacDFI}}(\omega)$, do not include the Sagnac effect. 
\section{Gravitational wave signal of neutron DFI}
\label{Gravitational wave signal of neutron DFI}
\subsection{Gravitational wave signal in a Sagnac-type neutron DFI}
Let us consider the Klein-Gordon equation for a particle with mass $m$,

\begin{align}
\label{K-G}
    \left( -\frac{1}{c^2}\frac{\partial^2}{\partial t^2} + \nabla^2 -\frac{m^2 c^2}{\hbar^2}\right) \phi(t,\bm{x}) = 0.
\end{align}
The wave function at time $t$ and position $\bm{x}$ with wavenumber vector $\bm{k}$ and angular frequency of neutrons $\Omega$ is

\begin{align}
\label{wave0}
    \phi_0(t,\bm{x}) = e^{i(\Omega t -\bm{k}\cdot \bm{x})}.
\end{align}
Due to the effect of GWs, Equation\,\ref{wave0} becomes

\begin{equation}
\label{wave}
    \phi(t,\bm{x}) = \phi_0(t,\bm{x})\left[1+i\phi^{\mathrm{gw}}((t,\bm{x})\right].\\
\end{equation}
In the TT gauge, if the GWs propagate through flat spacetime, we can write

\begin{equation}
\label{metric}
    ds^2 = -(cdt)^2 + \left[ \eta_{ab}+h_{ab}(t,\bm{x})\right]dx^{a} dx^{b}  ~(a, b = 1, 2, 3).
\end{equation}
Here, $\eta_{ab}$ is a metric of a flat space. In this coordinate system, from Equation\,\ref{K-G} we obtain

\begin{align}
   \frac{1}{c^2}\left( \Omega + \frac{\partial \phi^{\mathrm{gw}}}{\partial t} \right)^2 -k_a k_b \left(1-h^{ab}\right)^2 &=0.
\end{align}
The leading order term in $h^{ab}$ is

\begin{align}
   \frac{\partial \phi^{\mathrm{gw}}}{\partial t} &= -\frac{h^{ab}k_a k_b c^2}{2\Omega}\\
   &\approx -\frac{h^{ab}k_a k_b \hbar}{2m}.
\end{align}
Here, assuming that the wavenumber vector is sufficiently small, we can make the following approximation,

\begin{equation}
    \frac{\Omega ^2}{c^2} \approx \frac{m^2 c^2}{\hbar^2}.
\end{equation}

As mentioned in Section \ref{Configuration of Sagnac-type neutron DFI}, $n_i$ propagate through the interferometer on the trajectories represented by Equations\,1-13. Here we reset the time, that is, we set the time to be $t_0$ when the $n_i$ cross the surface of of BS A. The wavenumber vectors of the neutrons during propagation are given by

\begin{align}
    \bm{k}_{i\mathrm{AC}}(t) &= \frac{m}{\hbar}(-\frac{1}{\sqrt{2}}v_{i}^{\mathrm{h}}, -\frac{1}{\sqrt{2}}v_{i}^{\mathrm{h}}, v_{i}^{\mathrm{v}} - g(t-t_0)),\\
    \bm{k}_{i\mathrm{AD}}(t) &= \frac{m}{\hbar}(\frac{1}{\sqrt{2}}v_{i}^{\mathrm{h}}, -\frac{1}{\sqrt{2}}v_{i} ^{\mathrm{h}}, v_{i} ^{\mathrm{v}} - g(t-t_0))~,\\
    \bm{k}_{i\mathrm{CB}}(t) &= \frac{m}{\hbar}(\frac{1}{\sqrt{2}}v_{i} ^{\mathrm{h}}, -\frac{1}{\sqrt{2}}v_{i} ^{\mathrm{h}}, v_{i} ^{\mathrm{v}} - g(t-t_0))~,\\
    \bm{k}_{i\mathrm{DB}}(t) &= \frac{m}{\hbar}(-\frac{1}{\sqrt{2}}v_{i}^{\mathrm{h}}, -\frac{1}{\sqrt{2}}v_{i} ^{\mathrm{h}}, v_{i}^ {\mathrm{v}} - g(t-t_0)),\\
    \bm{k}_{i\mathrm{BD}}(t) &= \frac{m}{\hbar}(\frac{1}{\sqrt{2}}v_{i} ^{\mathrm{h}},  \frac{1}{\sqrt{2}}v_{i} ^{\mathrm{h}}, -v_{i} ^{\mathrm{v}} - g(t-t_0-4T_i)),\\
    \bm{k}_{i\mathrm{BC}}(t) &=  \frac{m}{\hbar}(-\frac{1}{\sqrt{2}}v_{i} ^{\mathrm{h}},  \frac{1}{\sqrt{2}}v_{i} ^{\mathrm{h}}, -v_{i} ^{\mathrm{v}} - g(t-t_0-4T_i)),\\
    \bm{k}_{i\mathrm{DA}}(t) &=  \frac{m}{\hbar}(-\frac{1}{\sqrt{2}}v_{i} ^{\mathrm{h}},  \frac{1}{\sqrt{2}}v_{i} ^{\mathrm{h}}, -v_{i} ^{\mathrm{v}} - g(t-t_0-4T_i)), \mathrm{and}\\
    \bm{k}_{i\mathrm{CA}}(t) &=  \frac{m}{\hbar}(\frac{1}{\sqrt{2}}v_{i} ^{\mathrm{h}},  \frac{1}{\sqrt{2}}v_{i} ^{\mathrm{h}}, -v_{i} ^{\mathrm{v}} -g(t-t_0-4T_i)).
\end{align}
The phase change of $\phi^{\mathrm{gw}}_i$ caused by GWs is given by

\begin{equation}
    \dfrac{\partial \phi^{\mathrm{gw}}_i}{\partial t} \approx -\frac{h^{a b}k_{i a} k_{i b}}{2m} \hbar. 
\end{equation}
Therefore, the phase change induced by the GWs during propagation between points red D and B is

\begin{equation}
    \phi_{i\mathrm{DB}}^{\mathrm{gw}}(t) = -\frac{\hbar}{2m} \int_{t}^{t + T_i} h^{ab}(t',\bm{x}_{\mathrm{DB}}(t'))k_{i\mathrm{DB}a}(t') k_{i\mathrm{DB}b}(t') d t'.  
\end{equation}
Defining $H^{a b}(\omega)$ as the Fourier transform of the GWs at angular frequency $\omega$,

\begin{equation}
     h^{a b}(t,\bm{x}(t)) = \int H^{a b}(\omega)e^{-i\omega t} d\omega.
\end{equation}
If we expand $h^{a b}(t)$ at the position $\bm{x}(t)$, we obtain the second term of order $\bm{k}_{\mathrm{gw}}\cdot \bm{x}(t)$. ($\bm{k}_{\mathrm{gw}}$ is a GW wave vector.) However, this second term is negligible because the typical size of a Sagnac-type DFI is $x \sim 100\,\mathrm{m}$, and $k_{\mathrm{gw}}x \sim x/\lambda_{\mathrm{gw}}$ is much smaller for GWs at 1Hz \cite{ref14}. For simplicity, we ignore the $\bm{x}(t)$-dependence of $H^{a b}(\omega)$.
In this paper, we calculate the response of a Sagnac-type neutron DFI to cross-mode GWs traveling in the z-axis direction.
The Fourier transform of the GW signal received by a neutron with velocity $v_i$ is:

\begin{equation}
    \Phi_{i\mathrm{DB}}^{\mathrm{gw}}(\omega) \approx -\frac{\hbar}{2m} \{ P_0(\omega) k_{i\mathrm{DB}a}k_{i\mathrm{DB}b}H^{a b}(\omega) \},
\end{equation}
 where

\begin{align}
    P_0(\omega) &= -\frac{i}{\omega}\left(1-e^{-i\omega T_i}\right).
\end{align}
When GWs of amplitude $h^{a b}$, polarization angle $\psi$, and angle of incidence $(\theta, \varphi)$ arrive, the corresponding rotation matrix is

\begin{equation}
 \mathcal{R} = \begin{pmatrix}
  \cos{\varphi} & \sin{\varphi} & 0 \\
 -\sin{\varphi} & \cos{\varphi} & 0 \\
  0          &          0 & 1 
\end{pmatrix}
\begin{pmatrix}
  \cos{\theta} & 0 & -\sin{\theta} \\
  0            & 1 & 0             \\
  \sin{\theta} & 0 & \cos{\theta} 
\end{pmatrix}
\begin{pmatrix}
  \cos{\psi} & \sin{\psi} & 0 \\
  -\sin{\psi}& \cos{\psi} & 0 \\
  0          &            0 & 1 
\end{pmatrix}
,
\end{equation}
and the GW strain $h^{'}_{a b}$ is

\begin{equation}
    h^{'}_{a b} = \mathcal{R}_{a p} \mathcal{R}_{b q} h_{p q} = (\mathcal{R} h \mathcal{R}^T)_{a b}.
\end{equation}
Similarly, the GW signals received by the clockwise- and counterclockwise-propagating neutrons are given by

\begin{align}
\label{phi_ir_gw}
    \phi_{i}^{\mathrm{c},\mathrm{gw}}(t) = \phi_{\mathrm{AD}}^{\mathrm{gw}}(t-T_i) + \phi_{\mathrm{DB}}^{\mathrm{gw}}(t) + \phi_{\mathrm{BC}}^{\mathrm{gw}}(t+T_i) + \phi_{\mathrm{CA}}^{\mathrm{gw}}(t+2T_i),\\
\label{phi_il_gw}
    \phi_{i}^{\mathrm{cc},\mathrm{gw}}(t) = \phi_{\mathrm{AC}}^{\mathrm{gw}}(t-T_i) + \phi_{\mathrm{CB}}^{\mathrm{gw}}(t) + \phi_{\mathrm{BD}}^{\mathrm{gw}}(t+T_i) + \phi_{\mathrm{DA}}^{\mathrm{gw}}(t+2T_i).
\end{align}
The Fourier transforms of Equation\,\ref{phi_ir_gw} and \ref{phi_il_gw} are

\begin{align}
   \Phi_{i}^{\mathrm{c},\mathrm{gw}}(\omega) = \Phi_{\mathrm{AD}}^{\mathrm{gw}}(\omega) e^{i\omega T_i} + \Phi_{\mathrm{DB}}^{\mathrm{gw}}(\omega) + \Phi_{\mathrm{B C}}^{\mathrm{gw}}(\omega) e^{-i\omega T_i} + \Phi_{\mathrm{CA}}^{\mathrm{gw}}(\omega) e^{-2i\omega T_i} \mathrm{and}\\
    \Phi_{i}^{\mathrm{cc},\mathrm{gw}}(\omega) = \Phi_{\mathrm{AC}}^{\mathrm{gw}}(\omega) e^{i\omega T_i} + \Phi_{\mathrm{CB}}^{\mathrm{gw}}(\omega) + \Phi_{\mathrm{B D}}^{\mathrm{gw}}(\omega) e^{-i\omega T_i} + \Phi_{\mathrm{DA}}^{\mathrm{gw}}(\omega) e^{-2i\omega T_i}.
\end{align}
For the DFI scheme, detection timing is crucial. In practice, the timing of neutron detection is affected by clock noise $\tau$. Clock noise in terms of neutrons’ phase is the cumulative effect of clock deviations at each mirror or BS due to reflections. We define the phase clock noise $\phi^{\mathrm{c}, \mathrm{clock}}_{i}(t)$ and $\phi^{\mathrm{cc}, \mathrm{clock}}_{i}(t)$ and their Fourier transforms $\Phi^{\mathrm{cc}, \mathrm{clock}}_{i}(\omega)$ and $\Phi^{\mathrm{cc}, \mathrm{clock}}_{i}(\omega)$ as

\begin{align}
    \phi^{\mathrm{c}, \mathrm{clock}}_{i}(t) = \phi^{\mathrm{cc}, \mathrm{clock}}_{i}(t) = \frac{m c^2}{\hbar} \left( \tau_{\mathrm{A}}(t+2T_i) - \tau_{\mathrm{A}}(t-2T_i)\right),\\
    \Phi^{\mathrm{c}, \mathrm{clock}}_{i}(\omega) =  \Phi^{\mathrm{cc}, \mathrm{clock}}_{i}(\omega) = \frac{m c^2}{\hbar} \left( e^{-2\omega T_i}\tau_{\mathrm{A}}(\omega) - e^{2\omega T_i}\tau_{\mathrm{A}}(\omega)\right),
\end{align}
The phase of the $n_i$ are given by

\begin{align}
    \Phi_i^{\mathrm{c}}(\omega) = \Phi_i^{\mathrm{c}, \mathrm{gw}}(\omega) + \Phi_i^{\mathrm{c}, \mathrm{dis}}(\omega) +  \Phi_i^{\mathrm{c}, \mathrm{clock}}(\omega),\\
    \Phi_i^{\mathrm{cc}}(\omega) = \Phi_i^{\mathrm{cc}, \mathrm{gw}}(\omega) + \Phi_i^{\mathrm{cc}, \mathrm{dis}}(\omega) +  \Phi_i^{\mathrm{cc}, \mathrm{clock}}(\omega).
\end{align}
The GW signal that the $n_i$ receive during propagation between A and C is given by
  
\begin{align}
\label{Phi_gw_DB}
    \Phi^{\mathrm{gw}}_{i \mathrm{DB}} &\approx -\frac{\hbar}{2m} \{ P_0(\omega) k_{\mathrm{DB}_a}k_{\mathrm{DB}_b}H^{a b}(\omega) \},\\ \nonumber
    &=i h^{\mathrm{gw}} \frac{m}{2\hbar \omega} (v_{i} ^{\mathrm{h}})^2 (1-e^{-i\omega T_i}),
\end{align}
where $h^{\mathrm{gw}}$ is the GW amplitude. 
Because of the difference in neutron propagation direction, GW signals that neutrons receive during propagation along each side of the Sagnac-type neutron DFI are expressed as

\begin{align}
\label{GW signals in each side}
    \Phi^{\mathrm{gw}}_{i \mathrm{AD}} = -\Phi^{\mathrm{gw}}_{i \mathrm{DB}}, \quad  \Phi^{\mathrm{gw}}_{i \mathrm{BC}} = -\Phi^{\mathrm{gw}}_{i \mathrm{DB}}, \quad  \Phi^{\mathrm{gw}}_{i \mathrm{CA}} = \Phi^{\mathrm{gw}}_{i \mathrm{DB}}.
\end{align}
Substituting Equation\,\ref{Phi_gw_DB} and \ref{GW signals in each side} into Equation\,\ref{phi_ir_gw}, $\Phi^{\mathrm{c},\mathrm{gw}}_{i}(\omega)$ can be written explicitly as

\begin{align}
    \Phi^{\mathrm{c},\mathrm{gw}}_{i}(\omega) &=\Phi^{\mathrm{gw}}_{i \mathrm{DB}}\ (-e^{i\omega T_1} + 1 - e^{-i\omega T_1} + e^{-2i\omega T_1}) \nonumber\\
    & = \Phi^{\mathrm{gw}}_{i \mathrm{DB}} [-(e^{i\omega T_i} + e^{-i\omega T_i}) + e^{-i\omega T_i}(e^{i\omega T_i} + e^{-i\omega T_i})] \nonumber\\
    &=\Phi^{\mathrm{gw}}_{i \mathrm{DB}} [2(-1+e^{-i\omega T_i})\cos(\omega T_i)]\nonumber\\
    &=i h^{\mathrm{gw}} \frac{m}{\hbar \omega} (v_{i} ^{\mathrm{h}})^2 (1-e^{-i\omega T_i})(-1+e^{-i\omega T_i})\cos{\omega T_i},\nonumber\\
        &=-i h^{\mathrm{gw}} \frac{m}{\hbar \omega} (v_{i} ^{\mathrm{h}})^2  e^{-i\omega T_i}  (e^{i\frac{\omega T_i}{2}}-e^{-i\frac{\omega T_i}{2}})^2\cos{\omega T_i},\nonumber\\
    &=-i h^{\mathrm{gw}} \frac{m}{\hbar \omega} (v_{i} ^{\mathrm{h}})^2 e^{-i\omega T_i}(2i \sin{\frac{\omega T_i}{2}})^2 \cos{\omega T_i}\nonumber\\
    &=4ih^{\mathrm{gw}} \frac{m}{\hbar \omega} (v_{i} ^{\mathrm{h}})^2 e^{-i\omega T_i}(1-\cos{\omega T_i})\cos{\omega T_i}.\label{Phi_gw}
\end{align}
The sign of the GW signal from the z direction received by the neutrons depends on the velocity of the neutrons in the xy-plane. The sign of the GW signal received by neutrons traveling along the same path in the opposite direction is opposite. Thus, $\Phi^{\mathrm{c},\mathrm{gw}}_{i}(\omega) = - \Phi^{\mathrm{cc},\mathrm{gw}}_{i}(\omega)$.\\
By combining signals for displacement noise cancellation in $\Phi_{i}^{\mathrm{c},\mathrm{g w}}(\omega)$ and $\Phi_{i}^{\mathrm{cc},\mathrm{g w}}(\omega)$, the GW signal $\Phi_{\mathrm{SagnacDFI}}^{\mathrm{gw}}$ contained in $\Phi_{\mathrm{SagnacDFI}}$ is calculated as

\begin{align}
    \Phi^{\mathrm{gw}}_{i}(\omega) = \Phi^{\mathrm{c},\mathrm{gw}}_{i}(\omega) - \Phi^{\mathrm{cc},\mathrm{gw}}_{i}(\omega) = 2\Phi^{\mathrm{c},\mathrm{gw}}_{i}(\omega).
\end{align}
Thus,

\begin{align}
    \Phi^{\mathrm{gw}}_{12}(\omega) &= \alpha_1 \frac{\sqrt{2}c}{v_1} \Phi^{\mathrm{gw}}_{1}(\omega) - \alpha_2 \frac{\sqrt{2}c}{v_2} \Phi^{\mathrm{gw}}_{2}(\omega), \quad \Phi^{\mathrm{gw}}_{34}(\omega) = \alpha_3 \frac{\sqrt{2}}c{v_3} \Phi^{\mathrm{gw}}_{3}(\omega) - \alpha_4 \frac{\sqrt{2}}c{v_4} \Phi^{\mathrm{gw}}_{4}(\omega),
\end{align}
and

\begin{align}
\begin{split}
    \Phi^{\mathrm{gw}}_{\mathrm{SagnacDFI}}(\omega) = \beta_1 \Phi^{\mathrm{gw}}_{12}(\omega) -\beta_2\Phi^{\mathrm{gw}}_{34}(\omega).
\end{split}
\end{align}

\subsection{Gravitational wave signal of a single, two-velocity neutron DFI}
Similar to the GW signals in the Sagnac-type neutron DFI, the GW signals in the single, two-velocity neutron DFI $\Phi_{\mathrm{SingleDFI}}^{\mathrm{gw}}$, are obtained as follows:

\begin{align}
    \Phi_{\mathrm{SingleDFI}}^{\mathrm{gw}} = ih^{\mathrm{gw}}\frac{m}{\hbar \omega}[\gamma_1(v_{1 \mathrm{SingleDFI}}^{\mathrm{h}})^2(1-e^{-i\omega T_1})^2 - \gamma_2(v_{2 \mathrm{SingleDFI}}^{\mathrm{h}})^2(1-e^{-i\omega T_2})^2],
\end{align}
where

\begin{align}
    \gamma_1 = \frac{\sqrt{2}c\sin{\omega T_2}}{v_{1 \mathrm{SingleDFI}}^{\mathrm{h}}} \mathrm{and} \quad \gamma_2 = \frac{\sqrt{2}c\sin{\omega T_1}}{v_{2 \mathrm{SingleDFI}}^{\mathrm{h}}}.
\end{align}

\section{Sensitivity}
\label{Sensitivity}
DFI signal, as its name implies, is free from displacement noise. Therefore, the sensitivity of a DFI to GW signals is limited by neutron shot noise. If the flux of the neutrons $n_i$ is $F_i$, then their shot noise $N_i$ is given by

\begin{equation}
    N_i = \frac{1}{\sqrt{F_i}}.
\end{equation}
In a neutron DFI, a signal without any displacement noise is obtained by combining signals from neutrons with four velocities. Since $N_1, N_2, N_3$, and $N_4$ are all independent of each other, the shot noise $N_{\mathrm{SagnacDFI}}$ after DFI combination is given by

\begin{equation}
    N_{\mathrm{SagnacDFI}} = \sqrt{\left(\alpha_1 \beta_1 \frac{\sqrt{2}c}{v_1^{\mathrm{h}}} N_1\right)^2 + \left(\alpha_2 \beta_1 \frac{\sqrt{2}c}{v_2^{\mathrm{h}}}N_2\right)^2 + \left(\alpha_3 \beta_2 \frac{\sqrt{2}c}{v_3^{\mathrm{h}}}N_3\right)^2 + \left(\alpha_4 \beta_2 \frac{\sqrt{2}c}{v_4^{\mathrm{h}}}N_4\right)^2}.
\end{equation}
Similarly, the shot noise of a single, two-velocity neutron DFI is given by \cite{ref15}

\begin{align}
    N_{\mathrm{Single DFI}} = \sqrt{2(\gamma_1 N_1)^2 + 2(\gamma_2 N_2)^2}.
\end{align}
Note that the factor of 2 in $\gamma_1 N_1$ and $\gamma_2 N_2$ comes from the fact that a single, two-velocity neutron DFI uses two neutron beams with velocities $v_1$ and $v_2$.
In the GW detection, the GW signal is expressed in the unit of $/ \sqrt{\mathrm{Hz}}$ as,

\begin{equation}
    S_{\mathrm{DFI}}(\omega) = |\Phi_{\mathrm{DFI}}^{\mathrm{gw}}|\omega^{1/2}.
\end{equation}
Setting $S_{\mathrm{DFI}}/N_{\mathrm{DFI}}$ = 1, we obtain the amplitude spectral density of shot noise limited sensitivity to GW amplitude $h_{\mathrm{n}}$ of the Sagnac-type neutron DFI in $/ \sqrt{\mathrm{Hz}}$.

\begin{align}
   h_{\mathrm{n}}(\omega) &\equiv H_{\mathrm{n}}(\omega)\omega^{1/2}\nonumber \\
   &=\frac{N_{\mathrm{DFI}}(\omega)}{\Phi^{\mathrm{norm}}_{\mathrm{DFI}}(\omega)},
\end{align}
where $\Phi^{\mathrm{norm}}_{\mathrm{DFI}}(\omega)$ is the GW signal normalized by the GW amplitude $h^{\mathrm{gw}}$,

\begin{align}
   \Phi^{\mathrm{norm}}_{\mathrm{DFI}}(\omega) \equiv \frac{S_{\mathrm{DFI}}(\omega)}{h^{\mathrm{gw}}}.
\end{align} 
In this paper, we optimize the sensitivity by varying the velocity of neutrons incident on the DFI.
We have calculated the sensitivity for various combinations of neutron velocities in the range of $v_i = 75\,\mathrm{m/s}$ to $100\,\mathrm{m/s}$, where similar fluxes are expected. Note that the neutron velocities were varied at $1\,\mathrm{m/s}$ intervals, and we assume a constant neutron flux at all velocities ($F_i = 10^{6}\,/\mathrm{s}$).
Table\,\ref{Neutron Velocity} shows the most sensitive combinations of neutron velocities.

\begin{table}[hbtp]
  \caption{Neutron Velocity}
  \label{Neutron Velocity}
  \centering
  \begin{tabular}{|c||c|c|c|}
    \hline
    i  & $v_i\,\mathrm{m/s}$ & $v_{i}^{\mathrm{h}}\,\mathrm{m/s}$ & $v_{i}^{\mathrm{v}}\,\mathrm{m/s}$ \\
    \hline
　    1 & 93.31  & 87.00  & 33.74  \\
　    2 & 75.00  & 69.00 & 29.39 \\
　    3 & 77.04  & 77.00  & 2.59  \\
　    4 & 100.00 & 100.00 & 0  \\
    \hline
    \end{tabular}
\end{table}
Let us compare the sensitivity of a Sagnac-type neutron DFI and a single, two-velocity neutron DFI with the same interferometer size ($L$ = $100\,\mathrm{m}$), four-velocity neutron sources, and the same neutron flux. These sensitivities are optimized for the incident neutron velocities.
The sensitivity of the single, two-velocity neutron DFI is calculated with $v_{1 \mathrm{SingleDFI}} = 75\,\mathrm{m/s}$ and $v_{2 \mathrm{SingleDFI}} = 100\,\mathrm{m/s}$.

\begin{figure}[H]
    \centering
    \includegraphics[clip,width=10.0cm]{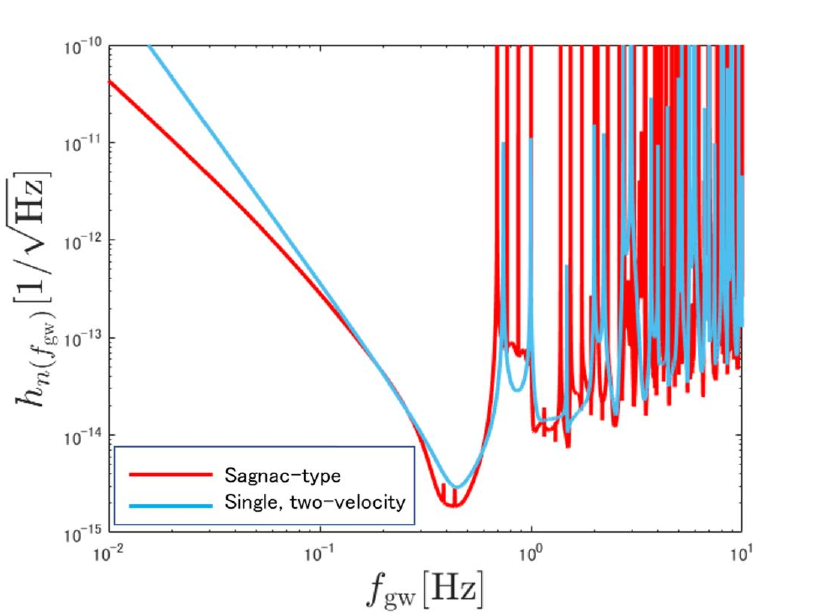}
    \caption{Sensitivities of a Sagnac-type neutron DFI and a single two-velocity neutron DFI.}
    \label{Sagnac_vs_Single}
\end{figure}
Figure\,\ref{Sagnac_vs_Single} shows that the sensitivity of the Sagnac-type neutron DFI is higher than that of the single, two-velocity neutron DFI around $0.4 \mathrm{Hz}$. 
This is indicated by the dip in the Sagnac-type curve being lower than that of the single, two-velocity curve at that frequency. The signal-to-noise ratio (SNR) is defined as

\begin{align}
    (\mathrm{SNR})^2 \equiv 4\int^{f_{\mathrm{max}}}_{f_{\mathrm{min}}} \frac{|h^{\mathrm{gw}}(f)|^2}{h_{\mathrm{n}}(f)^2} df,
\end{align}
where $f_{\mathrm{max}} = 10 \mathrm{Hz}$, $f_{\mathrm{min}} = 10^{-2} \mathrm{Hz}$.
The ratio of SNR  of the Sagnac-type neutron DFI to the single, two-velocity DFI, $\mathrm{S}_\mathrm{r}$, for GW signals from binary stars proportional to $f^{-\frac{7}{6}}$ \cite{ref8, ref17} with arbitrary magnitude is

\begin{equation}
\mathrm{S}_\mathrm{r} \equiv \frac{\mathrm{SNR}_{\mathrm{SagnacDFI}}}{\mathrm{SNR}_{\mathrm{SingleDFI}}} \approx 1.57.
\end{equation}

\section{Discussion}
\label{Discussion}
First, let us consider the amount of phase change experienced by $n_i$ as they propagate inside the interferometer due to GWs traveling in the z direction. Newtrons, $n_i$ traverse one side of the interferometer in time, $T_i$. When cross-mode GWs with a period of $2T_i$ pass the interferometer, the neutrons undergo the maximum phase shift. In the Sagnac-type neutron DFI, the neutrons are affected by GWs as they propagate through four sides of the interferometer. In a single, two-velocity neutron DFI the neutrons are affected by GWs only as they propagate through two sides of the interferometer. The neutrons in the Sagnac-type neutron DFI experience twice the phase change, $\Phi_i^{\mathrm{gw}}(\omega)$, compared to those in a single two-velocity neutron DFI.
Thus, the ratio of $\Phi_i^{\mathrm{gw}}(\omega)$ to $N_i$ in the Sagnac-type neutron DFI is twice that in a single two-velocity neutron DFI.
Figure\,\ref{Sagnac_vs_Single} shows the best sensitivity, $h_n$, of the Sagnac-type neutron DFI is $1.9 \times 10^{-15}\,/\sqrt{\mathrm{Hz}}$, and that of the single two-velocity neutron DFI is $2.9 \times 10^{-15}\,/\sqrt{\mathrm{Hz}}$. These magnitudes differ only by a factor of about 1.5, due to a difference in signal processing.
The frequency (period) of the GWs that produce the largest phase change depends on the $v_i^{\mathrm{h}}$ of the neutrons used to calculate the DFI signal. Since several neutrons, $n_i$, (i = 1 to 4) are used, 
the DFI signal is calculated by subtracting the neutron signals with different velocities from each other, and a part of the GW signals is also subtracted.
The Sagnac-type neutron DFI with more subtractions of neutron signals loses more GW signals in the signal processing than a single, two-velocity neutron DFI. Thus, the best sensitivity $h_n$ of the Sagnac-type neutron DFI and that of the single, two-velocity neutron DFI differ only by a factor of about 1.6.\\
Next, let us consider the peaks that appear in the Sagnac-type neutron DFI sensitivity curve.
These peaks have two origins. The first is that neutrons traverse square orbits, and the second is the coefficients $\alpha_i$ and $\beta_i$ used in signal processing.
According to Equation\,\ref{Phi_gw}, $\Phi_i^{\mathrm{gw}} \propto (1-\cos{\omega T_i})\cos{\omega T_i} =0$ when $\omega = \frac{2n\pi}{T_i} \quad \mathrm{or} \quad \frac{\pi}{T_i}(n-\frac{1}{2}) (n = 1, 2, ...)$.
This means that the sums of the GW signals with periods of $T_i/n$ or  $2T_i/(n-\frac{1}{2})$ ($n = 1, 2, ...$) on the $n_i$ is zero. First, when $n_i$ propagate along one side of the interferometer, cross-mode GWs with a period of $T_i/n$ induce a positive (or negative) phase change and an equal and opposite phase change. Second, the $n_i$ are not sensitive to GWs with a period of $T_i/n$. Neutrons with $v_i^{\mathrm{h}}$ reflected from the mirrors bend their propagation direction by 90 degrees every $T_i$.
Cross-mode GWs with a period of $2T_i/(n-\frac{1}{2})$ give a constant phase change while the neutrons propagate along one side of the interferometer, and an equal and opposite phase change while they propagate along the opposite. Therefore, the round trip phase change induced by the cross-mode GWs with a periods of $2T_i/(n-\frac{1}{2})$ is zero. Since the neutron signals $\Phi_i(\omega)$ do not have a GW signal with a specific period (frequency), the DFI is not sensitive to GWs at that frequency.

Next, let us focus on the coefficients used for signal processing $\alpha_i$ and $\beta_i$.
When $\omega = \frac{n\pi}{T_2}$, $\alpha_1 \equiv \sin{\omega T_2} = 0$. When $\omega = \frac{n\pi}{T_3}, \frac{n\pi}{T_4}$ or $\cos{\omega T_3} - \cos{\omega T_4} = 0$, $\beta_1 \equiv \sin{2\omega T_3}\sin{\omega T_4} - \sin{2\omega T_4}\sin{\omega T_3} = 2\sin{\omega T_3}\sin{\omega T_4} (\cos{\omega T_3} - \cos{\omega T_4})= 0$.
To summarize, $\alpha_1 \beta_1 \Phi_1^{\mathrm{gw}} = 0$ when

\begin{align}
     \omega = \frac{2n\pi}{T_1}, \frac{\pi}{T_1}(n-\frac{1}{2}), \frac{n\pi}{T_2}, \frac{n\pi}{T_3}, \frac{n\pi}{T_4} (n = 1, 2, ...) \quad \mathrm{or} \quad \cos{\omega T_3} - \cos{\omega T_4} = 0.
\end{align}
Similarly, $\alpha_1 \beta_1 \Phi_2^{\mathrm{gw}} = 0$ when

\begin{align}
     \omega = \frac{2n\pi}{T_2}, \frac{\pi}{T_2}(n-\frac{1}{2}), \frac{n\pi}{T_1}, \frac{n\pi}{T_3}, \frac{n\pi}{T_4} (n = 1, 2, ...) \quad \mathrm{or} \quad \cos{\omega T_3} - \cos{\omega T_4} = 0.
\end{align}
Thus, $\beta_1(\alpha_1 \Phi_1^{\mathrm{gw}} - \alpha_2 \Phi_2^{\mathrm{gw}}) = 0$ when

\begin{align}
     \omega = \frac{2n\pi}{T_1}, \frac{2n\pi}{T_2}, \frac{n\pi}{T_3}, \frac{n\pi}{T_4} (n = 1, 2, ...) \quad \mathrm{or} \quad \cos{\omega T_3} - \cos{\omega T_4} = 0.
\end{align}
The lowest frequencies that satisfy each of these conditions derived from $\frac{1}{T_1}, \frac{1}{T_2}, \frac{1}{2T_3}, \frac{1}{2T_4},$ and $\cos{\omega T_3} - \cos{\omega T_4} = 0$ are respectively,

\begin{align}
\label{12=0}
    f = 0.870\,\mathrm{Hz}, 0.69\,\mathrm{Hz}, 0.385\,\mathrm{Hz}, 0.500\,\mathrm{Hz}, \quad \mathrm{and} \quad f\approx 0.435\,\mathrm{Hz}.
\end{align}
Similarly, $\beta_2(\alpha_3\Phi_3^{\mathrm{gw}} - \alpha_4 \Phi_4^{\mathrm{gw}}) = 0$ when

\begin{align}
     \omega = \frac{n\pi}{T_1}, \frac{n\pi}{T_2}, \frac{2n\pi}{T_3}, \frac{2n\pi}{T_4} (n = 1, 2, ...) \quad \mathrm{or} \quad \cos{\omega T_1} - \cos{\omega T_2} = 0.
\end{align}
The lowest frequencies that satisfy each of these conditions derived from $\frac{1}{2T_1}, \frac{1}{2T_2}, \frac{1}{T_3}, \frac{1}{T_4},$ and $\cos{\omega T_1} - \cos{\omega T_2} = 0$ are respectively,

\begin{align}
\label{34=0}
    f = 0.435\,\mathrm{Hz}, 0.345\,\mathrm{Hz} (=\frac{1}{2T_2}), 0.770\,\mathrm{Hz}, 1.000\,\mathrm{Hz}, \quad \mathrm{and} \quad f\approx 0.385\,\mathrm{Hz}.
\end{align}
According to Equation\,\ref{12=0} and \ref{34=0}, $\Phi_{\mathrm{SagnacDFI}}^{\mathrm{gw}} = 0$ when

\begin{align}
\label{DFI=01}
    f = \frac{n}{T_1}, \frac{n}{T_2}, \frac{n}{T_3}, \frac{n}{T_4}, (n = 1, 2, ...)
\end{align}
in addition,  $\Phi_{\mathrm{SagnacDFI}}^{\mathrm{gw}} \approx 0$ at the frequencies corresponding to ($\frac{1}{2T_1}$ and $\cos{\omega T_3} - \cos{\omega T_4} = 0$) and ($\frac{1}{2T_3}$ and $\cos{\omega T_1} - \cos{\omega T_2} = 0$), respectively,

\begin{align}
\label{DFI=02}
    f \approx 0.435\,\mathrm{Hz} \quad \mathrm{and} \quad f\approx 0.385\,\mathrm{Hz}.
\end{align}
Thus, the sensitivity curve of $\Phi_{\mathrm{DFI}}^{\mathrm{gw}}(\omega)$ in Figure\,\ref{Sagnac_vs_Single} has the peaks at the frequencies in Equation\,\ref{DFI=01} and \ref{DFI=02}.

\section{Conclusion}
\label{Conclusion}
In this paper, we propose a Sagnac-type neutron DFI and have confirmed analytically that DFI can be realized by injecting neutrons into a Sagnac-type neutron interferometer considering the influence of gravity. We have also shown that the sensitivity of the Sagnac-type neutron DFI is superior to that of the conventional single, two-velocity DFI.
When canceling displacement noise with multiple signals, the location and time at which the neutrons receive the displacement noise are important.
In order to cancel the displacement noise of all test masses with four neutron signals, we adjust the neutrons' velocities and injection angles to satisfy the following conditions:
1) The neutron trajectories are symmetrical about the y-axis.
2) The four neutron groups are incident on the interferometer at the same point on BS A.
3) Neutrons, $\mathrm{neutrons}_1$ and $\mathrm{neutrons}_2$ ($\mathrm{neutrons}_3$ and $\mathrm{neutrons}_4$) hit the same points on $\mathrm{C}_1$ and $\mathrm{C}_2$ ($\mathrm{D}_1$ and $\mathrm{D}_2$) in Figure\,\ref{Sagnac_path}, respectively.
Additionally, we optimize the incident neutron velocities to achieve the best sensitivity for both the Sagnac-type neutron DFI and the single, two-velocity neutron DFI.

One of the technical challenges we face in realizing a Sagnac-type neutron DFI is the neutron reflection angle. Currently, mirrors with a reflection angle of only a few degrees have been created using current technology. However, there is no fundamental principle that limits the reflection angle. Thus, it is possible to develop a mirror with a large reflection angle, such as 45 degrees. Other practical issue that must be considered when demonstrating neutron DFI experimentally is the accuracy and capability of the instrument. The sensitivity of neutron DFI is affected by factors such as the accuracy of the distance between test masses, the reflectivity of the mirrors, the flux of neutrons and the accuracy of the displacement noise cancellation. For example, neutron beams with velocity ranging from 75 m/s to 100 m/s are limited to a flux of about $10^6$ /s using the PF2 beamline at the ILL reactor in France. In the future, fluxes of up to $10^9$ /s can be anticipated using beams from the ESS accelerator neutron source, which is currently under construction. We anticipate that even higher fluxes will be possible in principle, although we will have to wait for future facilities to be built. The accuracy of noise cancellation depends on the precision of neutron velocity determination (this is frequency noise). Increasing the distance from the neutron source to the interferometer increases the required accuracy of velocity determination. If the velocity is determined with an accuracy of 0.1 \% or better, the sensitivity curve will not be affected as long as the mirror displacement is below $10^{-11} \,\mathrm{m/\sqrt{Hz}}$.
For these reasons, developing such a technique is challenging, but we believe it is feasible. A more detailed study of these technical issues will be the subject of future work.

\section*{Acknowledgments}
We would like to thank R. L. Savage for English editing. This work was supported by the Japan Society for the Promotion of Science (JSPS) KAKENHI Grant Number JP19K21875. A. N. is supported by JSPS KAKENHI Grant Nos. JP23K03408, JP23H00110, and JP23H04893.


\end{document}